\documentclass[journal]{IEEEtran}

\usepackage{amsmath,amsfonts}
\usepackage{algorithmic}
\usepackage{array}
\usepackage[caption=false,font=normalsize,labelfont=sf,textfont=sf]{subfig}
\usepackage{textcomp}
\usepackage{stfloats}
\usepackage{url}
\usepackage{verbatim}
\usepackage{graphicx}
\def\BibTeX{{\rm B\kern-.05em{\sc i\kern-.025em b}\kern-.08em
    T\kern-.1667em\lower.7ex\hbox{E}\kern-.125emX}}
\usepackage{balance}
\usepackage{cleveref}
\usepackage{enumitem}
\usepackage{siunitx}
\usepackage{comment}
\usepackage{cite}

\usepackage{tikz}
\usepackage{pgfplots}
\pgfplotsset{compat=1.17}
\crefname{section}{Section}{Sections}
\crefname{equation}{Equation}{Equations}
\crefname{figure}{Figure}{Figures}
\crefname{table}{Table}{Tables}
\crefname{lstlisting}{Listing}{Listings}
\crefname{appendix}{Appendix}{Appendices}
\setlist[itemize]{leftmargin=10pt,itemindent=0pt,topsep=2pt,partopsep=2.5pt,parsep=1pt,itemsep=1.5pt,listparindent=\parindent{}}
\definecolor{gray90}{gray}{0.9}
\begin{document}
\title{Exploiting Decompression Latency\\for Covert Channels in Inter-Line-Compressed LLCs}
\author{David K. Oh, Hiroshi Sasaki%
\thanks{\textcopyright~2026 IEEE. Personal use of this material is permitted. Permission from IEEE must be obtained for all other uses, in any current or future media, including reprinting/republishing this material for advertising or promotional purposes, creating new collective works, for resale or redistribution to servers or lists, or reuse of any copyrighted component of this work in other works. DOI: \protect\url{https://doi.org/10.1109/LCA.2026.3737156}.}}

\newcommand{\name}{Covert Channel on XOR Cache}

\maketitle

\begin{abstract}
The recently proposed XOR cache is an inter-line-compressed last-level cache (LLC) that leverages the data-inclusion relationship between the private caches and the LLC, compressing two cache lines into one by XORing them.
The architecture relies on the cache coherence protocol for data decompression.
In this paper, we demonstrate that this mechanism---specifically the latency asymmetry between a cache hit on an uncompressed vs. compressed line---introduces microarchitectural vulnerabilities.
Based on this observation, we propose a 
covert channel attack targeting the XOR cache.
A colluding sender controls the receiver's access latency by triggering decompression through targeted write requests to partner cache lines.
By exploiting the data-dependent compression behavior of the XOR cache, the sender and receiver establish the channel using pre-agreed data values.
The channel achieves higher bandwidth than the Prime+Probe baseline for two reasons:
first, each bit is encoded in the compression state of an individual line rather than the occupancy of a cache set, so a single set carries multiple bits;
second, each bit is resolved by manipulating coherence-protocol state rather than forcing shared-cache evictions, so it costs fewer LLC accesses and demand misses than Prime+Probe.
Full-system simulations
show a bandwidth of \SI{2.9}{Mbps} at an observed \SI{0.98}{\percent} bit-error rate (BER) over 50,000 transmitted bits, $13.1\times$ the bandwidth of Prime+Probe under the same sub-\SI{1}{\percent}-BER selection rule.
\end{abstract}

\section{Introduction}\label{sec:introduction}

\IEEEPARstart{T}{he} cache hierarchy is essential for reducing memory access latency and sustaining high system performance.
As the working sets of modern applications continue to grow, last-level caches (LLCs) consume an increasingly large fraction of processor die area and power.
To increase the effective capacity of the cache, various cache compression techniques have been studied~\cite{panXORCacheCatalyst2025a,BDI2012}.

The XOR cache~\cite{panXORCacheCatalyst2025a} is a recently proposed LLC compression architecture that exploits redundancy created by the inclusion relationship between private caches and the LLC. It compresses two cache lines, $A$ and $B$, into a single encoded line by storing their bitwise XOR, $A \oplus B$.
To recover compressed data, the architecture relies on the cache coherence protocol to obtain the corresponding partner line when needed.
This design introduces a structural latency asymmetry: an access to an uncompressed line is handled as a normal LLC hit, while an access to a compressed line may trigger remote recovery if the requester does not already hold the partner line in its private cache.
This recovery path introduces additional coherence transactions and multi-hop routing latency.  

While the XOR cache has been evaluated primarily as a capacity-enhancing architecture, the security implications of its coherence-dependent decompression path remain unexplored.
In this paper, we demonstrate that this latency asymmetry constitutes a microarchitectural vulnerability.
We propose the XOR cache covert channel (XOR-CC), in which a sender modulates the LLC compression state, causing a receiver to observe either a normal LLC hit or a remote recovery delay.
In our encoding, a sender write forces decompression and turns the receiver's subsequent access into a fast LLC hit, whereas sender inactivity leaves the line compressed and causes the receiver to incur slow remote recovery.

Unlike traditional conflict-based channels which evict the receiver's lines (e.g.,~Prime+Probe~\cite{liuLastLevelCacheSideChannel2015}), the proposed channel operates through protocol state manipulation.
This difference has two consequences giving the channel a bandwidth advantage.
First, because each bit is carried by the compression state of an individual line rather than the occupancy of a whole set, a single set conveys multiple bits where a conflict-based channel conveys only one.
Second, because it manipulates coherence state rather than forcing LLC evictions, it generates fewer LLC accesses and demand misses per transmitted bit.

A recent study Safecracker~\cite{tsaiSafecrackerLeakingSecrets2020} showed that compressed caches leak information through data-dependent cache compressibility. 
Safecracker targets intra-line compression, where the compressed size of a line depends on its data and is observed through eviction pressure on the data array.
The key difference is that we instead target inter-line compression, where the compression state determines which coherence path serves a request, so the bit is decoded from hit latency alone rather than from eviction-induced misses.

This paper makes the following contributions.
First, we identify coherence-dependent decompression latency in the XOR cache as a new timing-channel primitive.
Second, we show how an attacker deliberately triggers unXORing through targeted writes to partner lines---deterministically modulating the receiver's latency---and build a covert-channel protocol on it.
Third, we observe that because each bit is carried by the compression state of an individual line rather than the occupancy of a set, a single set conveys multiple bits where a conflict-based channel conveys one, giving the channel its bandwidth advantage.
Finally, we evaluate the proposed channel with full-system simulations, showing a $13.1\times$ bandwidth improvement over Prime+Probe at a comparable observed BER below \SI{1}{\percent}.

\section{Background}\label{sec:background}

\subsection{Cache Covert Channels}
Cache covert channels establish communication by exploiting access latency differences in shared caches, allowing logically isolated entities to exchange data.
Prime+Probe~\cite{liuLastLevelCacheSideChannel2015} is a foundational technique used to construct such channels.
The communication protocol proceeds in three steps.
First, the receiver fills (primes) a targeted cache set with its own memory lines. 
Second, to transmit a bit `1', the sender accesses its own memory mapped to the same cache set, thereby evicting the receiver's data; to transmit a bit `0', the sender remains idle.
Third, the receiver re-accesses (probes) its initial lines, where the resulting access latency (a cache hit or miss) allows the receiver to directly infer the transmitted bit.

\subsection{The XOR Cache}\label{subsec:xorcache}
The XOR cache is an LLC compression architecture supporting both inter- and intra-cache-line compression.
This paper focuses on its inter-cache-line compression mechanism, which leverages the data-inclusion property between private caches and the LLC.

The architecture performs a bitwise XOR operation on a cache line ($A$) with a partner line ($B$), storing the compressed result ($A \oplus B$) in a single LLC line.
The result can then undergo intra-line compression.
To improve the effectiveness of intra-line compression, XOR Cache preferentially pairs lines with similar contents.
When a line is inserted into the LLC, the cache derives a 7-bit sparse byte labeling (SBL) map value from its contents and uses it to find a previously inserted line with the same map value.
Lines with identical contents necessarily generate the same map value and are therefore candidates for pairing, subject to the cache’s address-pairing constraints.
Thus, assigning the same pre-agreed value to both lines allows the sender and receiver to reliably form XOR pairs required by the attack.

When a read targets a compressed line, the XOR cache consults the coherence protocol and routes the request through one of three forwarding paths---local recovery, direct forwarding, or remote recovery---depending on which caches hold the requested and partner lines.

These paths differ in latency.
An uncompressed LLC hit returns data directly from the data array, whereas remote recovery is the slowest: when the requester holds neither the partner line $A$ nor a sharer of $B$, the LLC forwards $A \oplus B$ to $A$'s sharer, which reconstructs $B$ by XOR and returns it.
This latency difference is the timing primitive we exploit.
Separately, a write that upgrades $A$ to the Modified state makes the compressed copy stale, forcing an immediate unXORing that splits the pair into independent lines.
\section{Covert Channel on XOR Cache}\label{sec:proposal}

\begin{figure}[t]
    \centering
    \includegraphics[width=1.0\columnwidth]{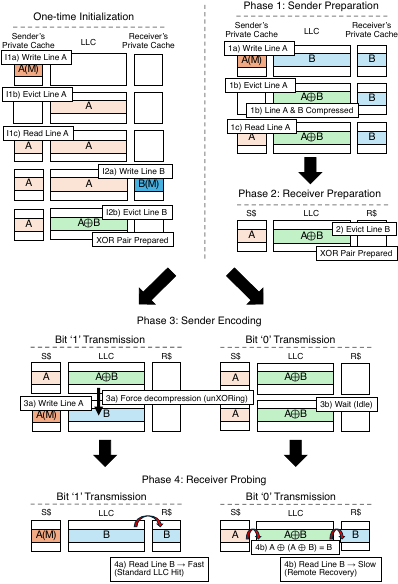}
    \caption{One-time initialization and repeated four-phase operation of XOR-CC, illustrated for one pair $(A_i,B_i)$. Initialization creates $A_i\oplus B_i$ with $A_i$ in the sender's private cache and $B_i$ absent from the receiver's private cache. Phase 1 uses the $B_i$ copy returned by the preceding probe to form the pair again and reloads $A_i$ through remote recovery; Phase 2 clean-evicts $B_i$; Phase 3 either writes $A_i$ for `1' or remains idle for `0'; and Phase 4 distinguishes a standalone LLC hit from remote recovery. The probe returns $B_i$ to the receiver for the next round.
    S\$ and R\$ denote the sender's and receiver's private caches, respectively.}
     \label{fig:attack_mechanism}
\end{figure}

\subsection{Threat Model}

We assume a standard covert-channel scenario in which two colluding processes execute on the same processor and share an LLC that employs the XOR cache.
Residing in distinct security domains, they cannot communicate through legitimate means and instead transmit information by modulating the shared LLC. 
Both processes run with unprivileged, user-level permissions; we assume no kernel compromise and no elevated privileges.
They use only ordinary loads and stores to drive the channel and a high-resolution timer (e.g., \texttt{rdtsc}) for the receiver to measure access latency.

\subsection{Attack Mechanism}\label{subsec:mechanism}

One pair of cache lines carries one bit (as opposed to one set carrying one bit in Prime+Probe), and thus multiple pairs can be transmitted per round in XOR-CC.
In our evaluated setting, four pairs are used since the L1D cache is four-way set associative.
The four pairs are mapped to the same L1D cache set and can be evicted from the private caches altogether when required by the protocol.
We select four address pairs $(A_i, B_i)$, $i \in \{0,\ldots,3\}$, that satisfy the XOR cache's address-pairing constraints, e.g., using reverse-engineered virtual-to-physical address mappings.
For each pair, the sender and receiver assign the same pre-agreed value to $A_i$ and $B_i$.

The protocol performs one-time initialization followed by repeated four-phase transmission rounds, as illustrated in \cref{fig:attack_mechanism}.

\noindent\textbf{One-time initialization.}
\begin{enumerate}
    \item[I1a)] The sender writes the assigned value to $A_i$.
    \item[I1b)] The sender evicts $A_i$ from its private caches.
    \item[I1c)] The sender reloads $A_i$ and keeps a private copy that the LLC can use when pairing $A_i$ with $B_i$.
    \item[I2a)] The receiver writes the corresponding value to $B_i$.
    \item[I2b)] The receiver evicts $B_i$. When $B_i$ reaches the LLC, the cache pairs it with $A_i$, whose private copy remains in the sender's cache.
\end{enumerate}
Initialization leaves $A_i$ in the sender's private cache, $B_i$ absent from the receiver's private caches, and $A_i\oplus B_i$ intact in the LLC. This is the transmission-ready state normally established at the end of Phase 2, so the first payload proceeds directly to Phase 3.

\noindent\textbf{Phase 1: Reset, pair formation, and sender reload.}
\begin{enumerate}
    \item[1a)] The receiver retains the $B_i$ copy returned by the preceding Phase 4 probe. The sender writes the assigned value to $A_i$. If the preceding bit left the pair intact, the write makes $A_i\oplus B_i$ stale and the cache first unpairs it; otherwise, the write preserves the value of $A_i$ while placing it in the Modified state.
    \item[1b)] The sender evicts $A_i$. When $A_i$ reaches the LLC, the cache pairs it with $B_i$, whose private copy remains in the receiver's cache.
    \item[1c)] The sender reloads $A_i$ through remote recovery. At the end of this step, the sender holds $A_i$, the receiver holds $B_i$, and $A_i\oplus B_i$ remains in the LLC.
\end{enumerate}

\noindent\textbf{Phase 2: Receiver preparation.}
\begin{enumerate}
    \item[2)] The sender remains idle while the receiver clean-evicts the $B_i$ copy supplied by the preceding probe. The pair remains intact, $A_i$ remains sender-private, and the receiver holds neither line. This completes the transmission-ready state.
\end{enumerate}

\noindent\textbf{Phase 3: Sender encoding.}
\begin{enumerate}
    \item[3a)] To transmit `1', the sender writes the assigned value to $A_i$. The write moves $A_i$ to the Modified state and makes $A_i\oplus B_i$ stale, so the cache unpairs it and stores $B_i$ as a standalone LLC line.
    \item[3b)] To transmit `0', the sender remains idle and the pair stays intact.
\end{enumerate}

\noindent\textbf{Phase 4: Receiver probing.}
\begin{enumerate}
    \item[4a)] After `1', the receiver times a load of standalone $B_i$, which returns as a fast LLC hit.
    \item[4b)] After `0', the receiver times a load of paired $B_i$, which triggers slower remote recovery. The LLC sends $A_i\oplus B_i$ to the sender's cache, where $B_i$ is reconstructed and forwarded to the receiver.
\end{enumerate}
In either case, the probe brings $B_i$ into the receiver's private cache, supplying the sharer-backed copy used to reform the pair in the next Phase 1.

\subsection{Timing Channel Distinguishability}\label{3.3}

The channel's viability rests on the latency gap between a standard LLC hit and remote recovery.
This gap arises from directory indirection (the LLC cannot supply the data directly and must forward the request to the sender holding the partner line), additional network-on-chip (NoC) hops, and cache-to-cache transfer overhead, including the remote XOR restoration.
Together these make the gap wide enough for the receiver to measure reliably with standard high-resolution timers (e.g., \texttt{rdtsc}).

\section{Evaluation}\label{sec:evaluation}

We evaluate XOR-CC against Prime+Probe in terms of communication bandwidth, bit-error rate and cache-traffic footprint.

\begin{table}[t]
\centering
\caption{Processor configuration.}
\label{tab:sys_config}
\resizebox{\columnwidth}{!}{
\begin{tabular}{l l}
\hline
\textbf{Component} & \textbf{Configuration Parameters} \\
\hline
\textbf{CPU} & 2 x86 cores, 4-issue superscalar OoO, \SI{3}{GHz} \\
\textbf{Guest OS} & Linux 4.19.83, Ubuntu 18.04 \\
\textbf{L1 I-Cache} & \SI{32}{KiB}, 4-way, 4-cycle, \SI{64}{B} line, Private \\
\textbf{L1 D-Cache} & \SI{32}{KiB}, 4-way, 4-cycle, \SI{64}{B} line, Private \\
\textbf{L2 Cache} & \SI{256}{KiB}, 8-way, 9-cycle, \SI{64}{B} line, Private \\
\textbf{LLC} & \SI{1}{MiB} per core, 16-way, 40-cycle, \SI{64}{B} line, Shared \\
\textbf{XOR Map Table} & 128 entries, Direct-mapped \\
\hline
\end{tabular}
}
\end{table}

\subsection{Experimental Methodology}\label{subsec:methodology}

We implement XOR-CC using the full-system gem5/Ruby model developed by the XOR Cache authors~\cite{panXORCacheCatalyst2025a,lowepower2020gem5}.
The model implements a three-level cache hierarchy and an XOR Cache-specific Ruby coherence protocol that captures coherence-controller execution and on-chip network timing.
\Cref{tab:sys_config} summarizes the system parameters.
We use the 7-bit SBL map function described in \Cref{subsec:xorcache}, as in the original XOR Cache paper.
The sender and receiver threads are pinned to distinct cores.

Before payload transmission, XOR-CC tests predefined patterns to identify four $(A_i,B_i)$ pairs that form reliably and exhibit the expected remote-recovery latency.
A timing diagnostic measured a median of 122 cycles for a standalone $B_i$ access and 139 cycles for remote recovery, a difference of 17 cycles.
These measurements include timestamp reads, private-cache miss handling, Ruby controller execution, and network traversal.
The four selected values and the receiver's latency threshold remain fixed for every payload, and the protocol accesses the four pairs sequentially within each phase.

For comparison, Prime+Probe follows prior work~\cite{liuLastLevelCacheSideChannel2015} on a standard Ruby configuration matched in core count, clock frequency, cache capacities, associativities and line sizes.
The receiver primes and probes two 16-way LLC sets using 16 congruent lines per set and decodes the bit from their aggregate latencies.
The sender transmits `0' or `1' by accessing 16 congruent lines in the corresponding set.

The sender and receiver synchronize each protocol phase using pre-agreed absolute TSC values. In our gem5 setup, both cores use a common TSC; each core-pinned thread waits for its assigned value using \texttt{rdtscp}, without shared-memory synchronization.
For each channel, we report the highest-bandwidth tested configuration whose observed pooled BER over 50,000 bits is below \SI{1}{\percent}.

For every tested timing configuration, we transmit ten independently seeded random payloads of \num{5000} bits each.
The same fixed seed list is reused across configurations and channels, yielding \num{50000} measured bits per configuration.
We report pooled BER as the total bit errors divided by \num{50000}.

\subsection{Channel Bandwidth}\label{4.2}

\Cref{tab:bw_ber} presents the results.
The phase-cycle entries give the scheduled duration of each consecutive phase in one transmission round.
Payload measurement begins after one-time initialization and pattern and threshold calibration; the reported bandwidth is calculated over this payload interval.

\begin{table}[t]
\centering
\caption{Highest-bandwidth tested configuration with observed pooled BER below \SI{1}{\percent}, over ten \num{5000}-bit runs. Phase cycles are listed in protocol order.}
\label{tab:bw_ber}
\resizebox{\columnwidth}{!}{
\begin{tabular}{l c r r r}
\hline
\textbf{Channel} & \textbf{Phase cycles} & \textbf{Errors} & \textbf{BER} & \textbf{Bandwidth (Kbps)} \\
\hline
XOR-CC & 2038/859/315/945 & 491 & 0.982\% & 2886.7 \\
Prime+Probe & 4999/2631/5994 & 455 & 0.910\% & 220.2 \\
\hline
\end{tabular}}
\end{table}

As shown in the table, XOR-CC reaches \SI{2886.7}{Kbps} ($\sim$\SI{2.9}{Mbps}) at an observed pooled BER of \SI{0.982}{\percent}, compared with \SI{220.2}{Kbps} at \SI{0.910}{\percent} for Prime+Probe. Under the same selection rule, XOR-CC provides $13.1\times$ the bandwidth of Prime+Probe.

\subsection{Cache Access and LLC-Miss Footprint}

\begin{table}[t]
\centering
\caption{Payload-only demand cache traffic per transmitted bit at the reported configurations, pooled over ten \num{5000}-bit runs.}
\label{tab:llc_traffic}
\resizebox{\columnwidth}{!}{
\begin{tabular}{l r r r r}
\hline
& \multicolumn{4}{c}{\textbf{Traffic per transmitted bit}} \\
\textbf{Channel}
& \textbf{L1D accesses}
& \textbf{L2C accesses}
& \textbf{LLC accesses}
& \textbf{LLC misses} \\
\hline
XOR-CC
& 175.2
& 10.2
& 6.5
& 0.380 \\
Prime+Probe
& 1197.9
& 91.5
& 82.7
& 52.8 \\
\hline
\end{tabular}}
\end{table}

\Cref{tab:llc_traffic} reports aggregate demand cache traffic per transmitted bit during payload transmission. XOR-CC generates 175.2, 10.2, and 6.5 demand accesses per bit at the L1D, L2C, and LLC, respectively, compared with 1197.9, 91.5, and 82.7 for Prime+Probe.
At the LLC, XOR-CC generates 0.4 demand misses per bit, compared with 52.8 for Prime+Probe.
XOR-CC produces fewer demand accesses at all three cache levels and fewer LLC demand misses per transmitted bit in the evaluated workload and system configuration.
This per-bit reduction in cache traffic, together with the fact that multiple bits (four in our environment) are transmitted per round, are the sources of XOR-CC's higher bandwidth.

\section{Discussion}

\textbf{Mitigations.}
First, XOR cache can block cross-domain pairing by requiring both lines of a pair to carry the same protection-domain identifier.
This blocks cross-domain pairing but requires additional domain metadata and forgoes cross-domain compression opportunities.
Second, it can eliminate the timing signal by delaying standalone LLC hits to match remote-recovery latency, following the constant-time principle, but this penalizes every LLC hit.
Third, a detection-based defense could count forced unXORing and remote-recovery events per core or protection domain and flag or throttle cores or domains that trigger them at unusually high rates. However, benign compressible workloads may also generate these events; evaluating whether reliable detection thresholds can distinguish such workloads from XOR-CC is left to future work.

\textbf{Dependence on the pairing policy.}
Our attack forms pairs by writing identical contents, which the XOR cache's similarity-based pair selection favors.
A vendor could instead implement inter-line compression alone, with a different pair-selection policy.
The timing primitive itself is unaffected, since the latency gap arises from the coherence path taken to recover a compressed line rather than from how partners are chosen; only the setup step changes, as the sender and receiver must construct line contents that the deployed policy pairs.
For a policy that pairs on any recoverable relation between two lines, this reduces to reverse-engineering that relation, and the remainder of the protocol carries over unchanged.
\section{Conclusions}\label{sec:conclusion}

While cache compression architectures such as the XOR cache improve resource efficiency, their management mechanism might inadvertently compromise system logical isolation.
In this paper, we proposed a covert channel attack that exploits the structural latency asymmetry between a standard LLC hit and the remote recovery triggered during data forwarding in the XOR cache; because the architecture relies on data-dependent compression, colluding processes can induce the required compressed states using pre-agreed values.
Full-system gem5 simulations show that XOR-CC reaches $13.1\times$ the bandwidth of Prime+Probe at a comparable observed BER below \SI{1}{\percent}.

\bibliographystyle{IEEEtran}
\bibliography{main}

\end{document}